\documentclass[prd,letterpaper,twocolumn,nofootinbib,longbibliography,floatfix,superscriptaddress]{revtex4-2}
\usepackage[utf8]{inputenc}
\usepackage{siunitx}
\usepackage{amssymb}
\usepackage{amsmath}
\usepackage{amsfonts}
\usepackage{graphicx}

\usepackage{booktabs}
\usepackage[colorlinks=true,allcolors=blue]{hyperref}
\usepackage{multirow, makecell}
\usepackage{listings}

\usepackage{floatrow}
\usepackage{ragged2e}

\DeclareSIUnit \s {\second}
\DeclareSIUnit \ns {\nano\second}
\DeclareSIUnit \mus {\micro\second}
\DeclareSIUnit \ms {\milli\second}
\DeclareSIUnit \MB {\mega\byte}
\DeclareSIUnit \GB {\giga\byte}
\DeclareSIUnit \TB {\tera\byte}
\DeclareSIUnit \PB {\peta\byte}
\DeclareSIUnit \Mbps {\mega\bit/\s}
\DeclareSIUnit \Gbps {\giga\bit/\s}
\DeclareSIUnit \Tbps {\tera\bit/\s}
\DeclareSIUnit \Pbps {\peta\bit/\s}
\DeclareSIUnit \kton {\kilo\tonne} % changed  back to kton
\DeclareSIUnit \kt {\kilo\tonne}
\DeclareSIUnit \Mt {\mega\tonne}
\DeclareSIUnit \eV {\electronvolt}
\DeclareSIUnit \keV {\kilo\electronvolt}
\DeclareSIUnit \MeV {\mega\electronvolt}
\DeclareSIUnit \GeV {\giga\electronvolt}
\DeclareSIUnit \TeV {\tera\electronvolt}
\DeclareSIUnit \PeV {\peta\electronvolt}
\DeclareSIUnit \EeV {\exa\electronvolt}
\DeclareSIUnit \m {\meter}
\DeclareSIUnit \cm {\centi\meter}
\DeclareSIUnit \in {\inchcommand}
\DeclareSIUnit \km {\kilo\meter}
\DeclareSIUnit \kV {\kilo\volt}
\DeclareSIUnit \kW {\kilo\watt}
\DeclareSIUnit \MW {\mega\watt}
\DeclareSIUnit \MHz {\mega\hertz}
\DeclareSIUnit \mrad {\milli\radian}
\DeclareSIUnit \year {years}
\DeclareSIUnit \POT {POT}
\DeclareSIUnit \sig {$\sigma$}
\DeclareSIUnit\parsec{pc}
\DeclareSIUnit\lightyear{ly}
\DeclareSIUnit\foot{ft}
\DeclareSIUnit\ft{ft}
\DeclareSIUnit \ppb{ppb}
\DeclareSIUnit \ppt{ppt}
\DeclareSIUnit \samples{S}
\DeclareSIUnit \pe{PE}

\newcommand\SigmaTwo{\SI{95.4}\percent}

\newcommand{\enu}{\E_\enu}

\begin{document}

\title{Closing the Neutrino ``BSM Gap'':\\Physics Potential of Atmospheric Through-Going Muons at DUNE}

\begin{abstract}
Many Beyond-Standard Model physics signatures are enhanced in high-energy neutrino interactions.
To explore these signatures, ultra-large Cherenkov detectors such as IceCube exploit event samples with charged current muon neutrino interactions $\gtrsim\SI{1}\TeV$.
Most of these interactions occur below the detector volume, and produce muons that enter the detector.
However, the large spacing between detectors leads to inefficiency for measuring muons with energies below or near the critical energy of $\SI{400}\GeV$.
In response, IceCube has built a densely instrumented region within the larger detector.
This provides large samples of well-reconstructed interactions that are contained within the densely instrumented region, extending up to energies of $\SI{\sim50}\GeV$.
This leaves a gap of relatively unexplored atmospheric-neutrino events with energies between $\SI{50}\GeV$ and $\SI{1}\TeV$ in the ultra-large detectors.
In this paper we point out that interesting Beyond Standard Model signatures may appear in this energy window, and that early running of the DUNE far detectors can give insight into new physics that may appear in this range.
\end{abstract}

\author{Austin Schneider}
\email{aschn@mit.edu}
\affiliation{Dept.~of Physics, Massachusetts Institute of Technology, Cambridge, MA 02139, USA}

\author{Barbara Skrzypek}
\email{bskrzypek@fas.harvard.edu}
\affiliation{Department of Physics \& Laboratory for Particle Physics and Cosmology, Harvard University, Cambridge, MA 02138, USA}

\author{Carlos A. Arg{\"u}elles}
\email{carguelles@fas.harvard.edu}
\affiliation{Department of Physics \& Laboratory for Particle Physics and Cosmology, Harvard University, Cambridge, MA 02138, USA}

\author{Janet M. Conrad}
\email{conrad@mit.edu}
\affiliation{Dept.~of Physics, Massachusetts Institute of Technology, Cambridge, MA 02139, USA}

\date{October 2020}

\maketitle

\section{Introduction}

Nature produces neutrino events at energies far higher than those accessible through current accelerator technology~\cite{Vitagliano:2019yzm}.
These high-energy neutrinos are predominantly produced in cosmic-ray air-showers, although a smaller proportion reach Earth from astrophysical sources~\cite{Abbasi:2020jmh}.
At very high energies, new physics process may affect the energy distribution of naturally produced $\nu_\mu$ and $\bar\nu_\mu$~\cite{Barenboim:2003jm, Beacom:2003nh, Arguelles:2015dca, Bustamante:2015waa, Shoemaker:2015qul, Arguelles:2019rbn}.
This, in turn, affects the energy distribution of $\mu^\pm$ produced in charged-current neutrino interactions.
These interactions can be observed through their products when they occur within a detector, but many more $\mu^\pm$ are produced in the material surrounding an experiment.
Due to the high rate of downward-going cosmic rays, it is difficult for even deep underground experiments to separate the muons of interest from background.
But, with the entire Earth as a shield, upward, through-going muons provide a clean channel of $\nu_\mu$ and $\bar\nu_\mu$ events for Beyond Standard Model (BSM) studies.

The giga-tonne scale IceCube South Pole Neutrino Observatory~\cite{Aartsen:2016nxy} has already demonstrated the fruitfulness of searching for new particles, new forces, and new space-time symmetries with an upward, through-going muon sample~\cite{TheIceCube:2016oqi,Aartsen:2017ibm,Aartsen:2017xtt,Aartsen:2020iky}.
However, the full IceCube detector is currently only able to measure the energy distribution of $\SI{>500}\GeV$ events, and above $\si\TeV$-scale muons~\cite{Aartsen:2020fwb} with high efficiency, due to the wide spacing of the photon detectors.
To improve sensitivity to lower energy events, IceCube has instrumented a small central region with more closely spaced photon detectors, called Deep Core~\cite{Collaboration:2011ym}, with an instrumented mass of $\SI{\sim30}{\mega\tonne}$.
This allows efficient reconstruction of fully contained $\nu_\mu$ events that occur within DeepCore, yielding a high statistics sample of well reconstructed events from $\SI{\sim5}\GeV$ to $\SI{\sim200}\GeV$, with most reconstructed events below $\SI{50}\GeV$~\cite{Aartsen:2017nmd,Aartsen:2019tjl}.
This leaves a gap between $\SI{\sim50}\GeV$ and $\SI{\sim1}\TeV$.
Below this region, most muons are contained, and their energy is inferred by measuring their track length.
Above this region, IceCube becomes efficient to radiated showers above the critical energy of $\sim\SI{1}\TeV$.
However, neither of these methods can be applied in the gap region, which leaves only the zenith information available for analysis~\cite{Barger:2011rc,Esmaili:2013vza}.
This is problematic because this gap-region may contain interesting Beyond-Standard-Model (BSM) signatures that cannot be found with zenith information alone.
In fact, it has already been shown that for certain parameters of non-standard interactions (NSI), the BSM signature may be entirely missed~\cite{Liao:2016reh,Esmaili:2018qzu,Dev:2019anc}.

Smaller, more highly instrumented Cherenkov detectors, such as the $\sim\SI{50}{\kilo\tonne}$ Super-K~\cite{Fukuda:2002uc}, do not provide sufficient information to fill the energy gap.
Super-K is able to select upward-through-going muons, by tagging their showering behavior, but is unable to reconstruct the muon energy.
The shower-tagged events have $\SI{>10}\GeV$ with a peak at $\si\TeV$-scale~\cite{Desai:2007ra}, partially covering the gap in the IceCube data.
In contrast to the muon energy, Super-K is able to accurately reconstruct the muon angle, enabling study of the atmospheric neutrino zenith distribution.
Super-K observes a deficit for neutrinos that traverse the Earth's core, which is feature of BSM models that have a signature in the gap region~\cite{Liao:2016reh,Esmaili:2018qzu,Dev:2019anc}.
This deficit could also be interpreted as a downward statistical fluctuation, and so Super-K cannot make a clear statement about these signatures without the energy spectrum information.

\begin{figure*}[tb]
\begin{center}
\includegraphics[width=0.9\textwidth]{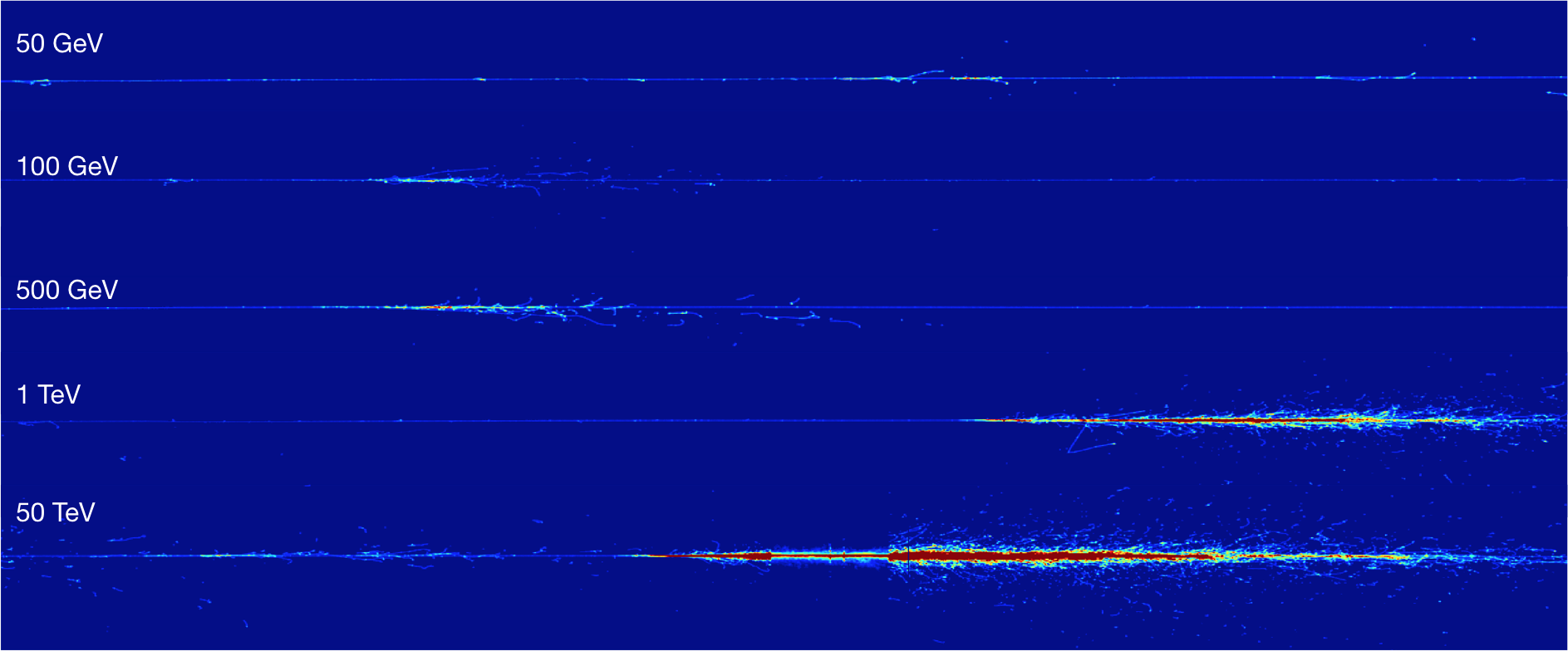}
\end{center}
\caption{\textbf{\textit{Illustration of the energy losses of muons in LarTPC.}}}
\label{fig:muon_illustration}
\end{figure*}

In this paper, we make the case that DUNE, a large underground neutrino experiment proposed to begin running by 2030~\cite{Abi:2020oxb}, can potentially explore BSM signatures in the gap region.
This paper touches on, but does not discuss in detail, approaches to measure the muon energy in the gap range.
Instead, the goal of this paper is to point out interesting physics that would be enabled by the development of algorithms to reconstruct the energy of $\SI{>50}\GeV$ muons traversing more than $\SI{2}\m$ in the DUNE far detector to the necessary precision.
To that end, the resolution necessary to do the proposed physics is discussed.
As physics motivation, we first explore a follow-up to the recent IceCube multi-year high-energy sterile neutrino search~\cite{Aartsen:2020iky}.
Secondly, we discuss signatures of Lorentz symmetry violation that DUNE may be sensitive to with information in this gap region.
Together these provide strong motivation for developing a dedicated reconstruction for through-going muons in DUNE.

\section{Energy Reconstruction in the Gap Region}

The fine-grained information available from the Liquid Argon Time Projection Chamber (LArTPC) technology of DUNE's $\SI{12}\m\times\SI{14}\m\times\SI{58}\m$ and $\SI{17(10)}\kt$ total (fiducial) volume far detectors~\cite{Abi:2020loh}, allows observation of radiative effects well below the critical energy.
For example, a $\SI{50}\GeV$ through-going muon traversing at least $\SI{2}\m$, which is the minimum path length we will consider, is expected to deposit $\SI{56}\MeV$ on average along the track.
Even if divided among several showers, this is much higher than the minimum photon energies that can be reconstructed in a LArTPC.
The ArgoNeut LArTPC, running in the Fermi National Accelerator Laboratory NuMI beam, has demonstrated reconstruction of $\si\MeV$-scale photons from de-excitation of the target argon nucleus, reconstructing a clear peak at $\SI{1.46}\MeV$, the first excited state of Argon~\cite{Acciarri:2018myr}.
The showering can be used to differentiate the muon energies, as illustrated in Fig.~\ref{fig:muon_illustration}, which shows the observed ionization when muons of different energies pass through $\SI{10}\m$ of liquid Argon.
Above $\SI{1}\TeV$, the electromagnetic showers are sufficiently long that they can partially exit the DUNE far detector modules.
This will limit the quality of the energy resolution for DUNE in the range that overlaps with IceCube.
But below $\SI{1}\TeV$, high-quality shower reconstruction can be expected.
Fortunately, the DUNE far detector modules are an appropriate size for resolving muons with energies in the gap range.

\begin{figure}[tb]
\begin{center}
\includegraphics[width=\linewidth]{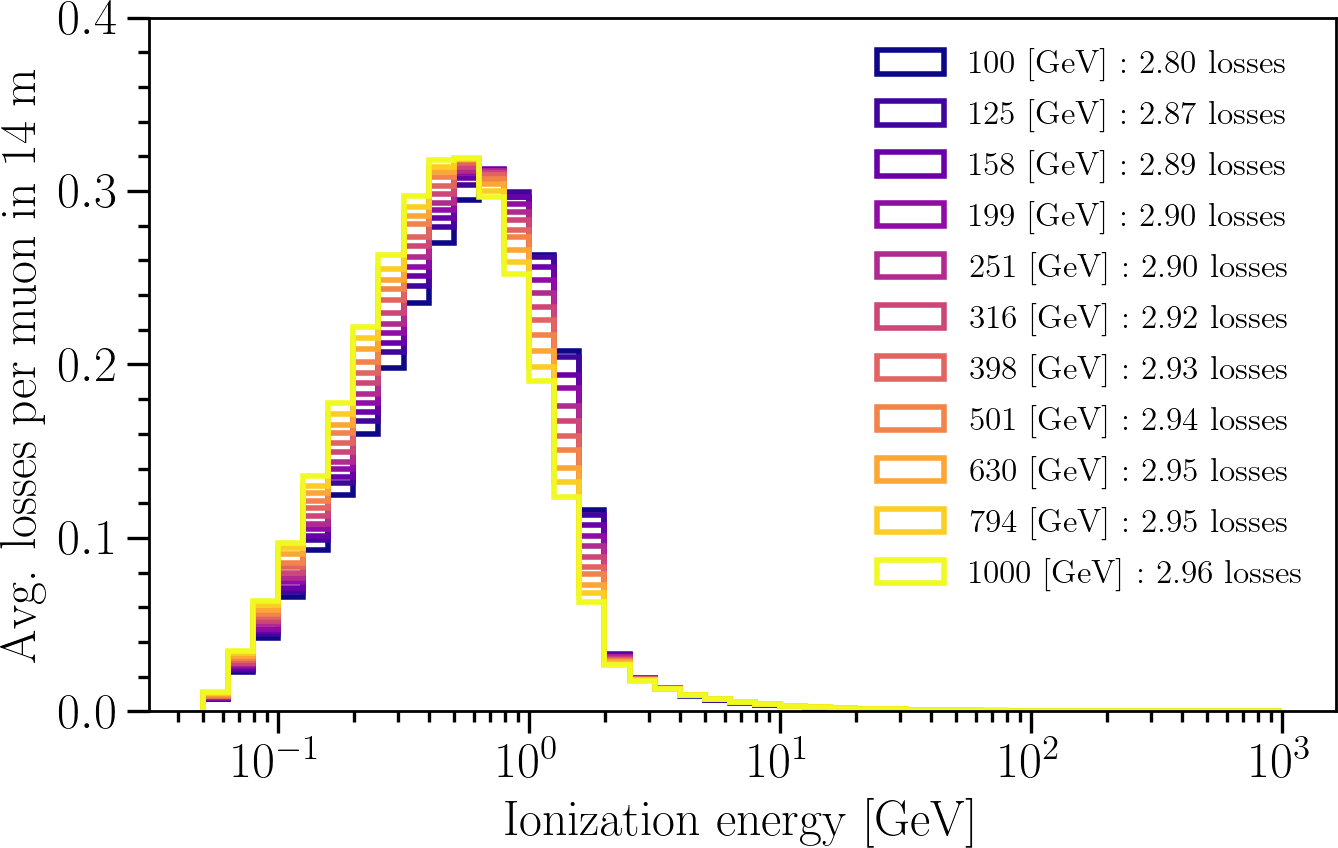}
\includegraphics[width=\linewidth]{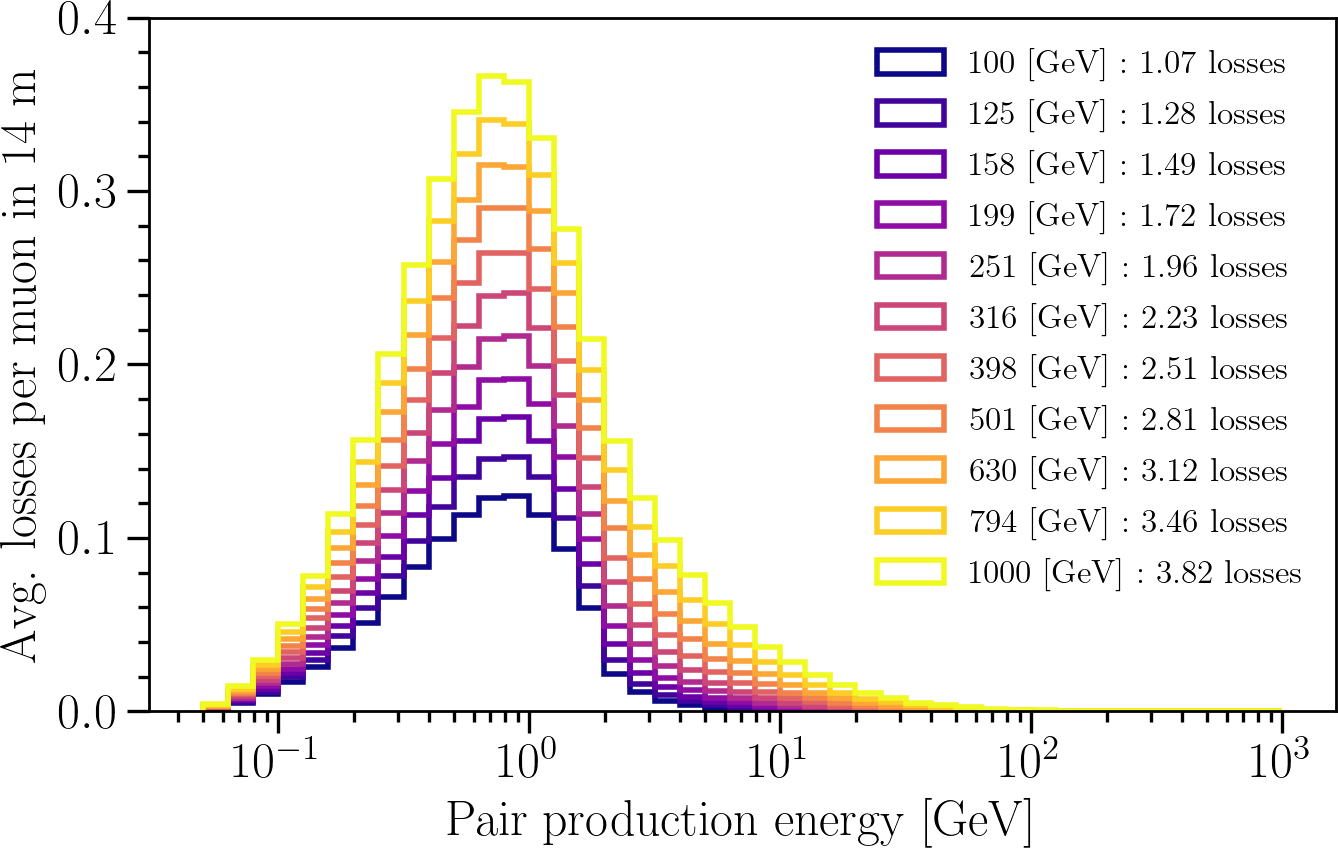}
\end{center}
\caption{\textbf{\textit{Muon energy loss expectation in Liquid Argon.}} The expected number of muon losses in $\SI{14}\m$ of Liquid Argon are shown as a function of the loss energy for different initial muon energies.}
\label{fig:muon_losses}
\end{figure}

In order to illustrate the potential of DUNE reconstruction, Fig.~\ref{fig:muon_losses} shows the distribution of energy losses for muons between $\SI{100}\GeV$ and $\SI{1}\TeV$ traversing the full $\SI{14}\m$ height of the DUNE far detector.
The top figure displays the energy distribution and expected number of ionization energy losses, which can produce $\delta$-rays.
The bottom figure shows the energy distribution and expected number of pair production energy losses, which can produce observable $e^+/e^-$ pairs.
Bremsstrahlung and photo-hadronic losses are also possible at these energies, but contribute less than $\SI{1}\percent$ of stochastic energy losses and total energy lost.
Simple counting of these energy losses can provide an energy resolution of $\sim\SI{40}\percent$, but more sophisticated reconstructions should be able to improve upon this.
Code will need to be developed to identify showers and $\delta$-rays in reconstructed events, but already the semantic segmentation code developed by MicroBooNE~\cite{Domine:2019zhm} shows promise of precise identification of both using machine learning.

\subsection{Modeling DUNE}

As part of the DUNE far-detector construction, two single-phase Liquid Argon Time Projection Chamber (LArTPC) modules are scheduled to begin operation before 2030, with the first module coming online one year before the other.
These modules will be located at the Homestake Mine $\SI{1478.27}\m$ level.
We assume that the liquid argon in each module occupies a $\SI{12}\meter\times\SI{14}\meter\times\SI{58.2}\meter$ region, and that the fiducial volume begins $\SI{10}\cm$ in from the walls.
Later, a third dual-phase LArTPC module, and a potential fourth detector will be installed.
However, in this work we consider only the two single-phase modules, with one available for 5 years of runtime and the other for 4 years.
We refer to this as the ``5 year'' scenario.

To estimate the neutrino rates and resulting rates of detectable events, we use publicly available software tools originally developed for use by the IceCube experiment.
The simulation consists of four steps:
\begin{enumerate}
    \item injection of neutrino interaction final-states in and around the detector,
    \item propagation of muons through surrounding rock,
    \item collection of muons intersecting the fiducial volume,
    \item and approximation of the detector response.
\end{enumerate}
LeptonInjector~\cite{LeptonInjectorGit,Abbasi:2020mwj} is used to inject Deep Inelastic Scattering (DIS) neutrino final states within the detector modules and in the surrounding material within a radius equal to the $\SI{99.9}\percent$ quantile of the muon range.
This injection scheme models the physical distribution of neutrinos that contribute to event signatures in the detector, primarily so that computation time is used efficiently in the remaining simulation steps.
Although it is possible to reweight all aspects of the injection, only the energy distribution, zenith distribution, and overall flux normalization are changed with respect to the injection.
For muon neutrino charged current DIS final states, a muon is produced at the interaction vertex.
These muons are passed to the PROPOSAL~\cite{dunsch_2020_1484180,koehne2013proposal,dunsch_2018_proposal_improvements} software package, which propagates high-energy muons and other charged particles through large distances of user-defined media.
For this study, we assume that far detector modules are surrounded by rock of density $\rho=\SI{2.65}{\g\per\cubic\cm}$.
Muons that intersect with or originate from the detector fiducial volume are recorded along with the properties of their parent neutrino.
These events are split into through-going events and other events, including starting, stopping, and fully contained events.
To approximate the reconstruction uncertainties, through-going and other events are given a ``reconstructed energy'' equal to the muon energy at entry to the fiducial volume multiplied by a log-normal distributed random smearing factor.
For through-going events the normal distribution has parameters $\mu=0,\sigma=0.2$ and for other events the parameters $\mu=0,\sigma=0.1$ are chosen.
The zenith angle of all events is smeared by a normal distribution with $\sigma=\SI{0.1}\degree$.

The final steps to obtaining physical data expectations involve weighting the simulation so that it resembles the atmospheric neutrino flux.
This process has three steps:
\begin{enumerate}
    \item removal of generation bias,
    \item choice of atmospheric neutrino flux at the Earth's surface,
    \item and propagation of the neutrino flux through the Earth assuming a BSM scenario.
\end{enumerate}
LeptonWeighter~\cite{LeptonInjectorGit,Abbasi:2020mwj} is the companion software to LeptonInjector, and is used to weight the events, removing the generation distribution biases introduced by LeptonInjector.
We model the atmospheric neutrino flux at the Earth's surface with a baseline atmospheric neutrino flux.
This is computed with MCEq~\cite{MCEq,Fedynitch:2015zma}, assuming the Hillas-Gaisser H3a~\cite{Gaisser:2013bla,Gaisser:2012zz,Hillas:2006ms} cosmic-ray model, and SIBYLL 2.3c~\cite{Riehn:2017mfm} hadronic interaction model.
Splines of this flux calculation in the nuflux~\cite{nufluxGit} repository are queried to obtain the flux for this analysis.
To model the propagation of neutrinos through the Earth, the nuSQuIDs~\cite{nusquids,Delgado:2014kpa} package is used, which models attenuation, oscillations, matter effects, BSM effects including sterile oscillations, and other relevant physical processes.

This analysis focuses on neutrino events with energy $\SI{>100}\GeV$; to this end, we aim to accept muon tracks with ``reconstructed'' energy $\SI{>100}\GeV$.
These can be identified because radiative effects are turning on at these energies, with $\SI{>1}\percent$ of the energy loss of a $\SI{>100}\GeV$ muon appearing as photon showers.
Energy reconstruction can utilize the count and energy distribution of photon showers, $\delta$-rays, and protons knocked out by fast neutrons.
Past experiments have successfully used only photon showers to isolate events in this energy range, as discussed in~\cite{Chikkatur:1997si}.
The use of $\delta$-rays and fast neutrons will be new information for energy reconstruction -- available because of the fine-grained information provided by LArTPC detector technology.
The average track energy of $\delta$-rays and showers will vary by $\SI{10}\percent$ between $\SI{100}\GeV$ and $\SI{1}\TeV$.
While this is a small effect, it will be well-measured, providing additional information to a machine learning reconstruction.
Lastly, the DUNE LArTPC will be instrumented with light collection which can also potentially be used~\cite{Ingles:2018}.
For the remainder of the discussion, we will assume that a $\SI{20}\percent$ energy resolution on through-going muons can be achieved, however a resolution of $\SI{50}\percent$ does not change our assessment of the sensitivity.

\begin{figure}[tb]
\begin{center}
\includegraphics[width=\linewidth]{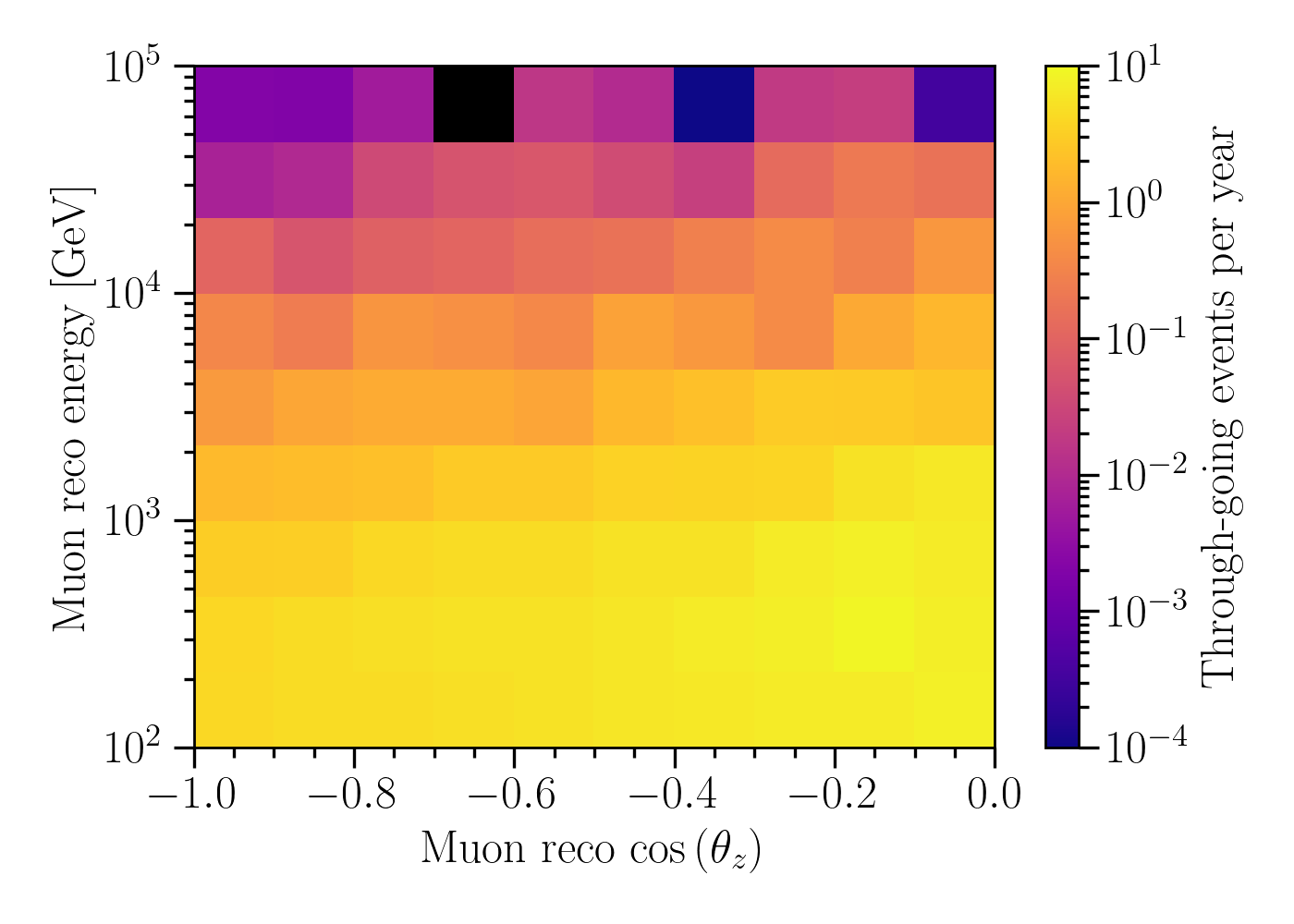}
\includegraphics[width=\linewidth]{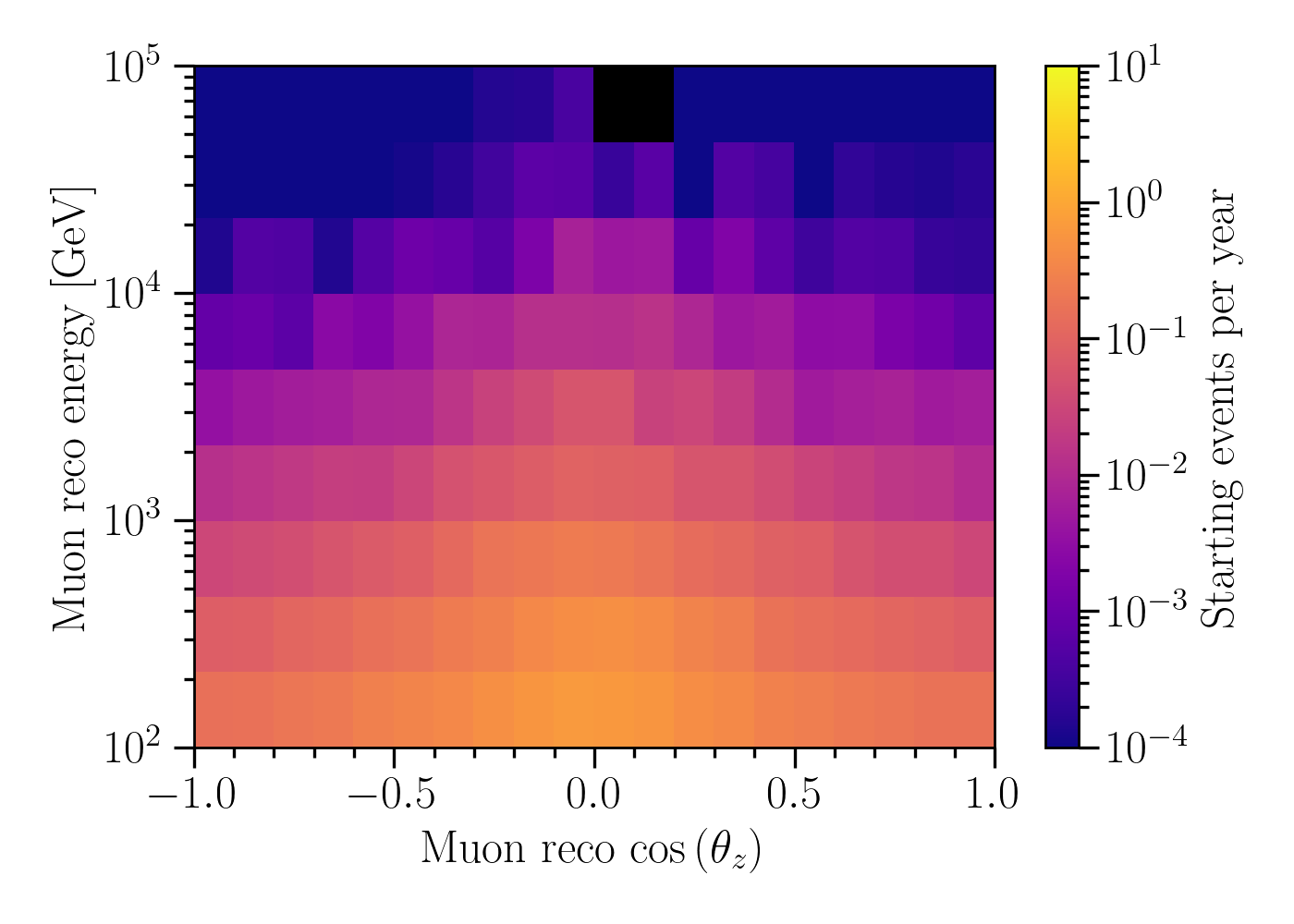}
\end{center}
\caption{\textbf{\textit{Observable distribution of through-going muon events.}} The distribution of through-going muon events in observable quantities is given for one module-year.}
\label{fig:through_going_distribution}
\end{figure}

To avoid cosmic-ray air-shower muon backgrounds we only use through-going muon events that come from below the horizon ($\cos(\theta_z) < 0$).
We expect that the contamination of downward-going cosmic ray muon background to the upward-going sample will be negligible, even at, or slightly above the horizon~\cite{Abe:2014wla}.
Also, reconstructing the direction of a high-energy cosmic-ray or muon in a LArTPC is straightforward by observing the direction of the $\delta$-rays produced along the track.
With this directional constraint we expect $\sim2070$ through-going events above $\SI{100}\GeV$ in the 5 year scenario.
The observable distribution of these events for one module-year is given in Fig.~\ref{fig:through_going_distribution}.

Although through-going muons comprise most of the observable events, we will still observe $\sim117$ muon neutrino charged-current events above $\SI{100}\GeV$ with an interaction vertex inside the fiducial volume in the five year scenario.
For these events we will be able to observe the hadronic shower produced in the neutrino interaction, providing us with more information about the neutrino energy.
DUNE has a large overburden compared to IceCube, and so the down-going cosmic-ray rate per square meter is two orders of magnitude less than IceCube.
The cosmic-ray muon flux at the $\SI{1478.27}\m$ level is predicted to be $\SI{4.4e-9}{\square\cm\per\s}$~\cite{Mei:2005gm}.
With this small background level, and the ability of the LArTPC to identify and reject down-going through-going cosmic-ray events, we can use contained vertex events from all directions.
The uncertainty on the cross section $>\SI{100}\GeV$ from accelerator-based neutrino experiments~\cite{Zyla:2020zbs, Tzanov:2005kr, Seligman:1997fe} is $\SI{2}\percent$.
With this small cross-section uncertainty, these events can constrain the conventional atmospheric neutrino flux normalization at the $\SI{10}\percent$ level, independent of the through-going sample.
The observable distribution of these ``starting'' events for one module-year is shown in the bottom panel of Fig.~\ref{fig:through_going_distribution}.

\subsection{DUNE Analysis Setup}
With the simulated event information and weighting procedure, we can now convert any oscillation hypothesis to an expected distribution of events in observable quantities.
Sensitivity to BSM scenarios can be determined by forming a binned likelihood and computing exclusion contours in the BSM parameter space assuming the average observed data for the null hypothesis.
This procedure is often referred to as computing the ``Asimov sensitivity''~\cite{Cowan:2010js,Asimov}.
In this case we split starting events and through-going events into two separate 2d-histograms with $40$ and $20$ $\cos\theta_z$ bins respectively, between $-1$ and $1$ for starting events and between $-1$ and $0$ for through-going events.
Both histograms have $30$ $\log E_\mu^\textrm{\tiny{reco}}$ bins between $\SI{100}\GeV$ and $\SI{100}\TeV$, bringing the total number of bins to $1800$.
Approximately $83000$ simulation events pass the cuts and are used to compute the expected event distribution.
To compare simulation to data, we used a Poisson-based binned-likelihood that accounts for simulation sample errors~\cite{Arguelles:2019izp} to avoid over-estimating the sensitivity.
We also introduce two dimensionless systematic parameters for the conventional atmospheric neutrino flux: the conventional normalization, $\Phi_\texttt{conv}$, and the cosmic-ray spectral index shift, $\Delta\gamma_\texttt{CR}$.
These modify the flux to account for uncertainties in the normalization and spectrum so that the conventional flux is
\begin{equation}
    \phi_\texttt{conv}(E_\nu, \theta_\nu^z)=\phi_\texttt{conv}^\textrm{nominal}(E_\nu, \theta_\nu^z)~\Phi_\texttt{conv}~\left(\frac{E}{\SI{500}\GeV}\right)^{-\Delta\gamma_\texttt{CR}},
\end{equation}
where $\phi_\texttt{conv}^\textrm{nominal}$ is the nominal conventional flux.
In the physics scenarios we explore, only the nominal conventional flux is modified by the BSM parameters.
We assume a $\sigma=0.05, \mu=1.0$ Gaussian prior on $\Phi_\texttt{conv}$ and a $\sigma=0.01, \mu=1.0$ prior on $\Delta\gamma_\texttt{CR}$.
These parameters are fit freely as nuisance parameters when computing the profile-likelihood.
The code for the DUNE simulation and statistical techniques used in this analysis is available in~\cite{DUNEAtmo}.

\section{BSM Physics Models}

\subsection{Sterile Neutrinos}

\begin{figure}[tb]
\begin{center}
\includegraphics[width=\linewidth]{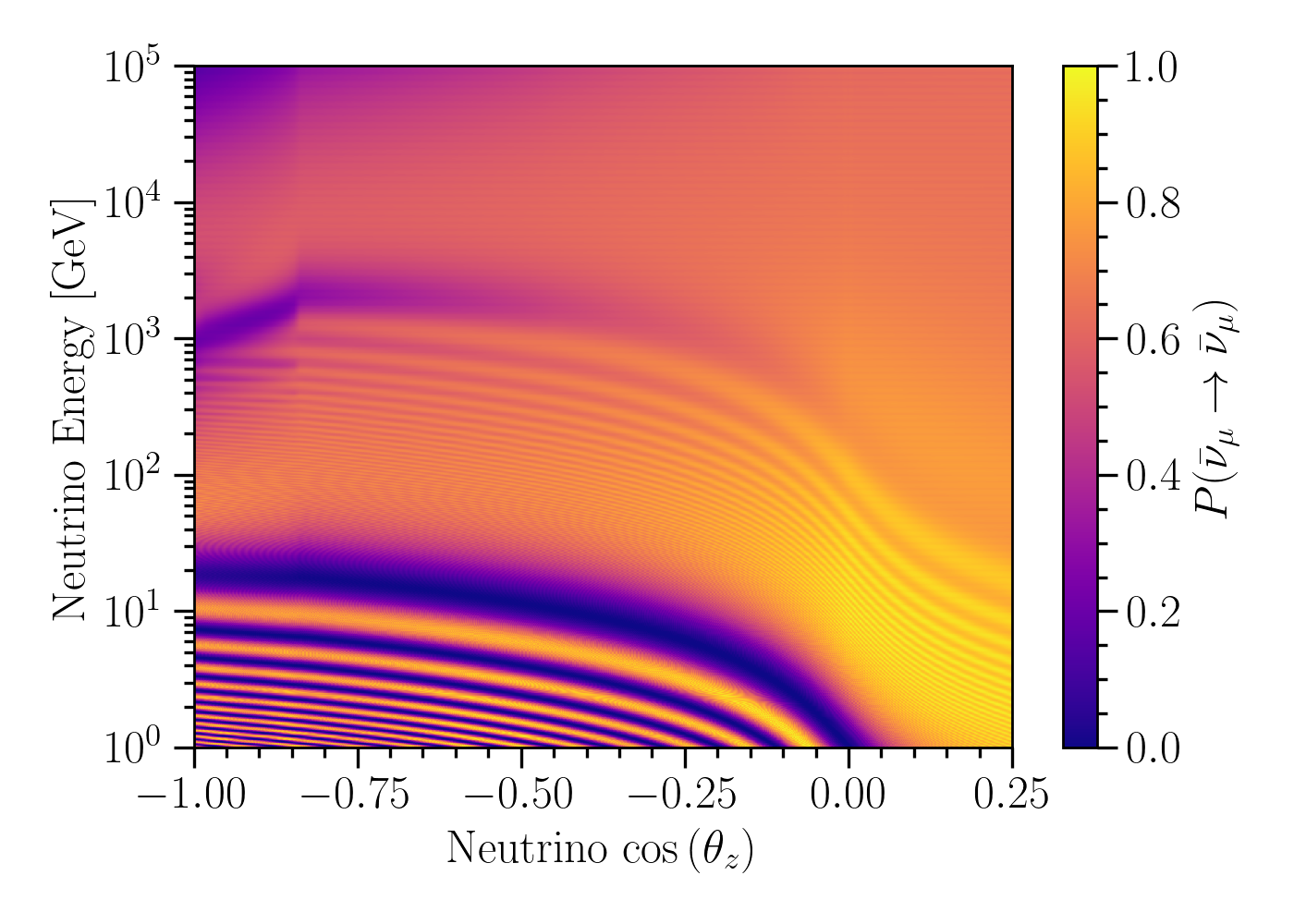}
\end{center}
\caption{\textbf{\textit{Sterile Neutrino Oscillogram.}} The transition probability of $\bar\nu_\mu\rightarrow\bar\nu_\mu$ is shown as a function of their energy and zenith angle for neutrinos passing through the Earth.
In this scenario a sterile neutrino is introduced, giving rise to a matter-enhanced resonance near $\SI{1}\TeV$.}
\label{fig:sterile_oscillogram}
\end{figure}

\begin{figure}[tb]
\begin{center}
\includegraphics[width=\linewidth]{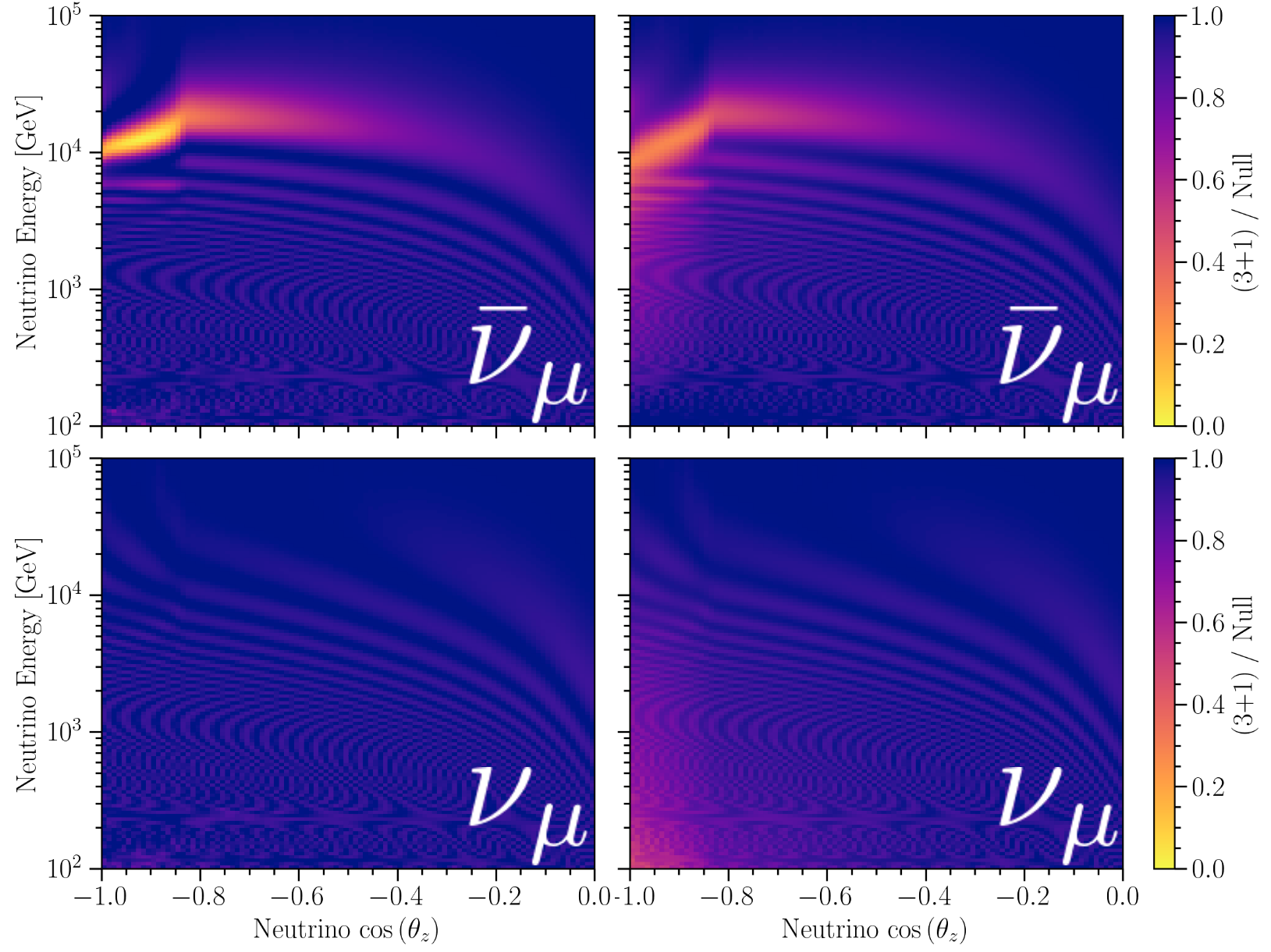}
\end{center}
\caption{\textbf{\textit{Sterile Neutrino Flux Ratio.}} The ratio between the expected atmospheric neutrino fluxes of two oscillation scenarios is shown in each panel.
The denominator in each case is the flux assuming only three neutrino states, all of which are active.
The numerator is the flux assuming a fourth sterile neutrino in addition to the three active neutrinos.
In the left panels $\theta_{34}$ is set to zero, and the right panels have non-zero $\theta_{34}$; all other mixing parameters are the same between panels.}
\label{fig:th34_oscillogram}
\end{figure}

Various anomalies in neutrino oscillation data samples are individually consistent with oscillations due to a light sterile neutrino~\cite{AguilarArevalo:2009xn,Cogswell:2018auu}.
However, as global-fits of neutrino data demonstrate~\cite{Gariazzo:2017fdh,Dentler:2018sju,Diaz:2019fwt, Moulai:2019gpi}, the null results from other experiments are not compatible with the simplest sterile-neutrino solutions~\cite{Armbruster:2002mp,Almazan:2018wln,Mahn:2011ea,Abe:2014gda,MINOS:2016viw,TheIceCube:2016oqi,Aartsen:2017bap,Adamson:2017uda,Albert:2018mnz}.
Options for addressing this ``tension'' include expanding to more complicated models~\cite{Liao:2016reh,Arguelles:2018mtc,Denton:2018dqq,Moulai:2019gpi} and searching for unidentified background sources.

The baseline new-physics model invoked to explain these anomalies introduces a new neutrino species with no standard model interactions but with mixing to the three active-flavors and is called ``3+1,'' known as a sterile neutrino.
This simple model suffers from ``tension'' when $\nu_\mu$ disappearance results are combined with $\nu_e$ disappearance and $\nu_\mu \rightarrow \nu_e$ appearance data.
The connection arises because the 3+1 model depends on four parameters: $\sin^2 2\theta_{ee}$, $\sin^2 2\theta_{\mu\mu}$, $\sin^2 2\theta_{\mu e}$, and $\Delta m^2$.
The first three are mixing angles measured in $\nu_e$ disappearance, $\nu_\mu$ disappearance, and $\nu_\mu \rightarrow \nu_e$ appearance, respectively.
Written in terms of the $4\times4$ flavor-mixing-matrix elements, one finds the mixing angles are not independent:
\begin{align}
\begin{split}
\sin^22\theta_{ee} ={}& 4(1-|U_{e4}|^2)|U_{e4}|^2; \\
\sin^22\theta_{\mu \mu } ={}& 4(1-|U_{\mu 4}|^2)|U_{\mu 4}|^2; \\
\sin^2 2\theta_{e\mu} ={}& 4|U_{e4}|^2 |U_{\mu4}|^2.
\end{split}
\end{align}
Also, the sterile mass squared splitting must be consistent for all three categories of data set.

A $\Delta \chi^2$ test comparing the scenarios indicates $>5\sigma$ improvement of a 3+1 model with respect to a 3 neutrino only model~\cite{Diaz:2019fwt}.
This large change arises because two $\nu_\mu$ disappearance experiments have $>\SI{90}\percent$ C.L. (but $<\SI{95}\percent$ C.L.) allowed regions in good agreement with the $\nu_e$ disappearance and appearance anomalies~\cite{Diaz:2019fwt}.
This would initially lead one to think 3+1 is an excellent explanation.
However, when one checks for consistency between these data samples with a Parametric Goodness of Fit test~\cite{Maltoni:2003cu}, a p-value of $3.7\times 10^{-6}$ is obtained~\cite{Diaz:2019fwt}, indicating a serious underlying problem with the model.
This poor consistency is driven by strict limits from some $\nu_\mu$ data sets.
These results will improve somewhat with the addition of the latest IceCube upward through-going muon-based search for $\nu_\mu$ disappearance, which has an allowed region in agreement with the anomalies at the $\SI{92}\percent$ C.L.~\cite{Aartsen:2020fwb}.
However, this result is sufficiently weak that the picture will remain murky.

Given this confusing situation, the options for interpreting the data are:
\begin{enumerate}
    \item all experiments with anomalies have separate systematic issues that, unfortunately, conspire to give very similar, but not identical oscillation parameters in a 3+1 fit,
    \item the 3+1 model is too simplistic and an improved BSM model is needed,
    \item or a combination of systematic issues and BSM physics is affecting the data.
\end{enumerate}

To resolve the current state of the field, input from new experiments will be needed, especially those with better controlled systematics and larger data samples.
DUNE seeks to do exactly this, but an analysis of accelerator neutrinos will come long after the detector has been constructed.
An upward through-going muon analysis with DUNE will allow us to address these three possibilities without waiting for the beam to come online.

For the (3+1) sterile neutrino model, the most relevant parameters for muon neutrino disappearance are $\Delta m^2$ and $\theta_{24}$.
These parameters control the location and shape of the matter resonance, which arises from the absence of a matter potential for the sterile neutrino state.
Figure~\ref{fig:sterile_oscillogram} shows the transition probability for $\bar\nu_\mu$ passing through the Earth.
In this case the matter resonance lies well above the region where oscillations from active neutrino mass splittings dominate.
The sterile mixing parameters have been constrained by the IceCube muon neutrino disappearance measurements.
However, IceCube's measurements and others set an important parameter, $\theta_{34}$, to zero.
Non-zero $\theta_{34}$ can smear the $\bar\nu_\mu$ matter resonance to lower energies, and cause additional disappearance for $\nu_\mu$ below the resonance energy.
Figure~\ref{fig:th34_oscillogram} demonstrates the effect of non-zero $\theta_{34}$ on the muon neutrino flux.
This modification to the oscillation signature can affect measurements of the matter resonance if not accounted for.
In the presence of true non-zero $\theta_{34}$ we would expect IceCube to measure a resonance at smaller $\Delta m^2$ than their current best-fit point.

\subsection{Lorentz Violation}

The observation of neutrino flavor oscillations by the Super-K and SNO experiments was one of the first indications that neutrinos exhibit behavior that is unaccounted for by the Standard Model.
Within the framework of the neutrino Standard Model ($\nu$SM), this phenomenon is attributed to a non-zero neutrino mass.
In this context, neutrinos are produced and detected in the flavor basis but propagate in the mass basis, oscillating between different flavor states.
Thus neutrinos are natural interferometers~\cite{Aartsen:2017ibm}, which makes them sensitive to tiny effects that accumulate along the neutrino propagation. 
Interferometric measurements have played an pivotal role in understanding the nature of vacuum and testing special relativity~\cite{michelson1887relative}.
Given the long baselines associated to neutrino oscillation measurements, it is not surprising that neutrino flavor morphing studies provide some of the stringent tests of Lorentz symmetry~\cite{Arguelles:2015dca, Aartsen:2017ibm}.

In order to study the sensitivity of DUNE to Lorentz symmetry violation, we consider the scenario where a Lorentz symmetry violating field permits space and can interact with neutrinos from their sources to the detector~\cite{Colladay:1998fq}.
Massive neutrino oscillations in vacuum are modelled by the following Hamiltonian:
\begin{equation}
    H_{\text{m}} = \frac{m^2}{2E} = \frac{1}{2E}UM^2U^{\dagger},~M^2 = \begin{pmatrix}
    m_1^2 & 0 & 0 \\
    0 & m_2^2 & 0 \\
    0 & 0 & m_3^2 \\
    \end{pmatrix}.
\end{equation}
In order to include the afored mentioned, Lorentz violation effects we include the following terms~\cite{Kostelecky:2003cr}:
\begin{equation}
    H \sim \frac{m^2}{2E} + \sum_{d\geq 3}p_{\mu}^{d-3}(a^{\mu(d)}-c^{\mu}{(d)}),
\end{equation}
where the zero component of the coefficients ($\mu = 0$) represents the isotropic component of the Lorentz-violating field, while the spatial components represent a direction-dependent field.
The isotropic component gives rise to modifications of the neutrino oscillation probability that are time independent, while the spatial component introduce time-varying neutrino flavor morphing amplitudes~\cite{Kostelecky:2003cr}.
In this work, we focus on the former case and study the effects of Lorentz symmetry violating operators on the angular and energy distribution.
These operators are classified as either $CPT$-odd ($a^{(d)}$) or $CPT$-even ($c^{(d)}$), and in the two-flavor basis, they can be expressed, in the isotropic case, as
\begin{equation}
    a^{(3)} = \begin{pmatrix}
    a_{\mu\mu}^{(3)} & a_{\mu\tau}^{(3)} \\
    a_{\mu\tau}^{{(3)}^{*}} & a_{\tau\tau}^{(3)} \\
    \end{pmatrix}.
\end{equation}

Without loss of generality, we can take these matrices to be traceless, whereby we are left with three independent parameters ($a_{\mu\mu}^{(3)},\text{Re}(a_{\mu\tau}^{(3)}),\text{Im}(a_{\mu\tau}^{(3)})$).
The off-diagonal terms dominate neutrino oscillations at high energies and are responsible for flavor-violation, while the diagonal terms contribute to the quantum Zeno effect, suppressing flavor changes.
We can quantify the strength of Lorentz violation using the expression $\rho_{\mu\tau}^{(d)} \equiv \sqrt{(a_{\mu\mu}^{(d)})^2+\text{Re}(a_{\mu\tau}^{(d)})^2+\text{Im}(a_{\mu\tau}^{(d)})^2}$~\cite{Aartsen:2017ibm}.
In terms of this quantity, we can label regions of the parameter space and map the exclusion region for tests of Lorentz-violation.
For example, near $a_{\mu\mu}^{(3)}/\rho_{\mu\tau}^{(3)} = -1$ and 1, Lorentz-violation is dominated by a large diagonal component, whereas at $a_{\mu\mu}^{(3)}/\rho_{\mu\tau}^{(3)} = 0$, Lorentz-violation exhibits maximal flavor violation.
\par
\begin{figure}[tb]
\begin{center}
\includegraphics[width=3.5in]{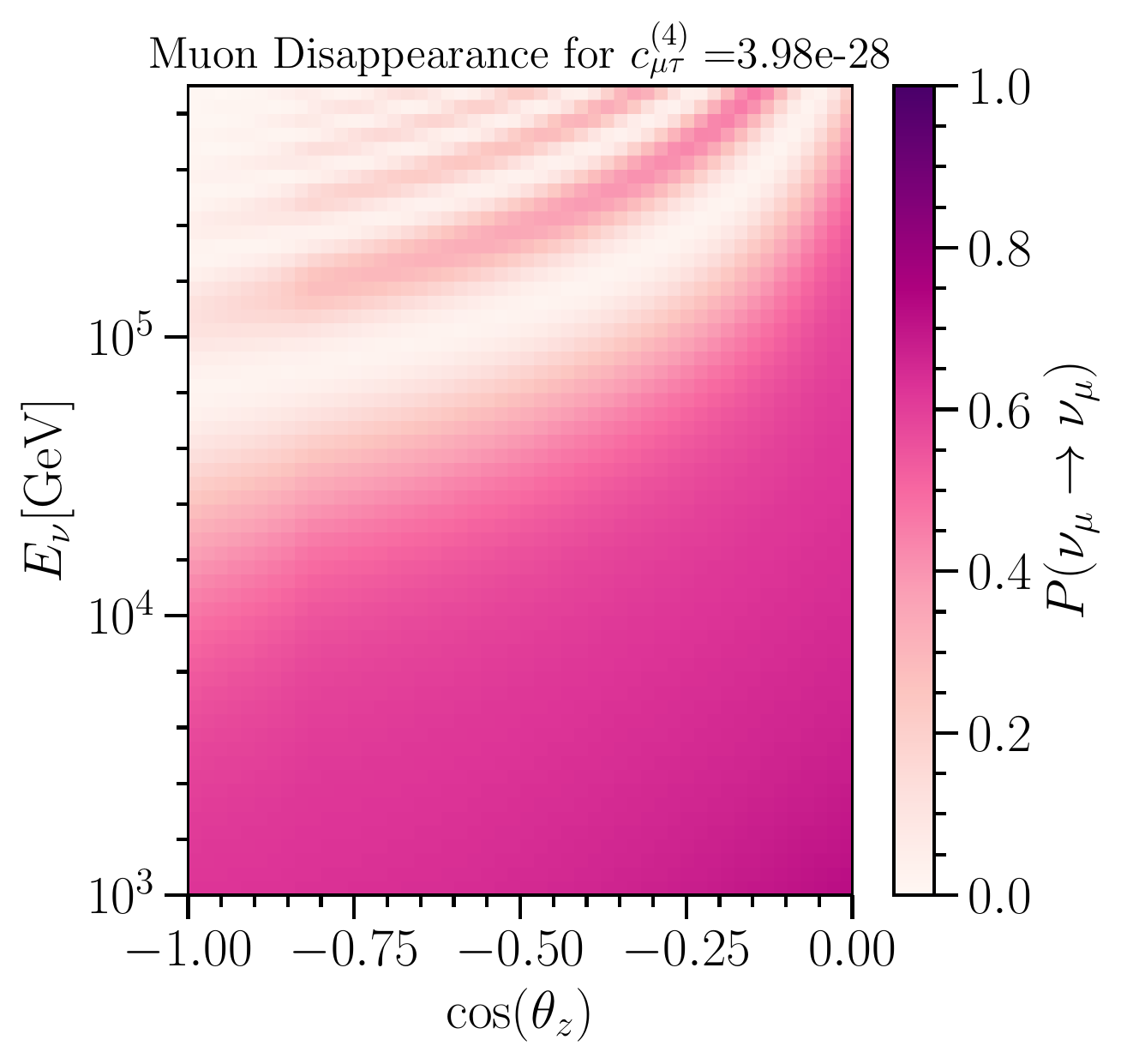}
\end{center}
\caption{\textbf{\textit{Lorentz Violation Oscillogram.}} Shown here is the survival probability for a muon neutrino of different energies and $\cos(\theta_z)$ undergoing Lorentz-violation effects from a non-zero value of the isotropic dimension-four operator. Here, we begin to observe oscillations induced by this operator at above $10$ TeV, which is where mass-induced oscillations for neutrinos traveling the distance of the diameter of the earth or less are heavily suppressed.}
\label{fig:lv_oscillogram}
\end{figure}

 The lower-dimensional operators, such as dimension-three and dimension-four, have been probed by terrestrial experiments using anthropogenic sources such as short-baseline accelerator neutrinos, long-baseline accelerator neutrinos, and reactor neutrinos~\cite{Auerbach:2005tq,AguilarArevalo:2011yi,Adamson:2008aa,Adamson:2010rn,Adamson:2012hp,Rebel:2013vc,Abe:2012gw,Diaz:2013iba,Abe:2017eot}, as well as with natural sources such as the solar or atmospheric neutrinos~\cite{Abbasi:2010kx,Abe:2014wla,Diaz:2016fqd}; for a summary of current constraints see~\cite{Kostelecky:2008ts}.
However, high-order terms are more difficult to constrain~\cite{Kostelecky:2011gq}, since they standout over the standard neutrino oscillation Hamiltonian only at high energies.
Currently, the best attainable limits, shown in Table~\ref{tbl:limits}, that we have on higher-dimensional operators such as dimensions-five, -six, and -seven, come from IceCube neutrino oscillation analyses of atmospheric neutrinos, owing to the long propagation lengths and high energies, Fig.~\ref{fig:lv_oscillogram} shows Lorentz-violation induced oscillations for a value of the diagonal dimension-four operator in terms of the $\nu_{\mu}$ transition probability.

However, it is important to note that the limits quoted above assume the maximum-flavor violating scenario, namely when the diagonal component dominates over the off-diagonal terms.
The scenarios in which the diagonal component dominates cannot be constrain by neutrino oscillations at high energies, since there are no standard model oscillations in this regime and the signature is the same as the standard model.
Thus, the strongest constraints on non-maximum-flavor-violating scenarios will be obtained for long-baseline experiments that observe neutrino oscillations at high energies.
To exemplify the complementarity between very-high-energy measurements, where standard neutrino oscillations are not present, to high-energy measurements, where muon-neutrino disappearance is present, we study in detail the dimension three operator in Sec.~\ref{sec:result}.

\begin{table}[t!]
\centering
\begin{tabular}{ c r l }
\toprule
    $(d)$ & Limit \\
    \midrule
    3 & $|\text{Re}(a_{\mu\tau}^{(3)})|,|\text{Im}(a_{\mu\tau}^{(3)})| <$ & $2.9\times 10^{-24}$ GeV\\
    4 & $|\text{Re}(c_{\mu\tau}^{(4)})|,|\text{Im}(c_{\mu\tau}^{(4)})| <$ & $3.9\times 10^{-28}$\\
    5 & $|\text{Re}(a_{\mu\tau}^{(5)})|,|\text{Im}(a_{\mu\tau}^{(5)})| <$ & $2.3\times 10^{-32}$ GeV $^{-1}$\\
    6 & $|\text{Re}(c_{\mu\tau}^{(6)})|,|\text{Im}(c_{\mu\tau}^{(6)})| <$ & $1.5\times 10^{-36}$ GeV $^{-1}$\\
\bottomrule
\end{tabular}
\caption{\textbf{\textit{Current limits on Lorentz violation.}}}
\label{tbl:limits}
\end{table}

\section{Results\label{sec:result}}

We can now take the simulated sample of through-going atmospheric-neutrinos and use it examine the sterile neutrino and Lorentz violation scenarios.

A significant number of events are present in the straight up-going region around $\SI{1}\TeV$, providing sensitivity to the sterile neutrino matter-enhanced-resonance.
However, many more events come from the horizon, where atmospheric neutrino production is peaked.
This also provides some sensitivity to the vacuum oscillations from the sterile neutrino mass splitting.

\begin{figure}[tb]
\begin{center}
\includegraphics[width=\linewidth]{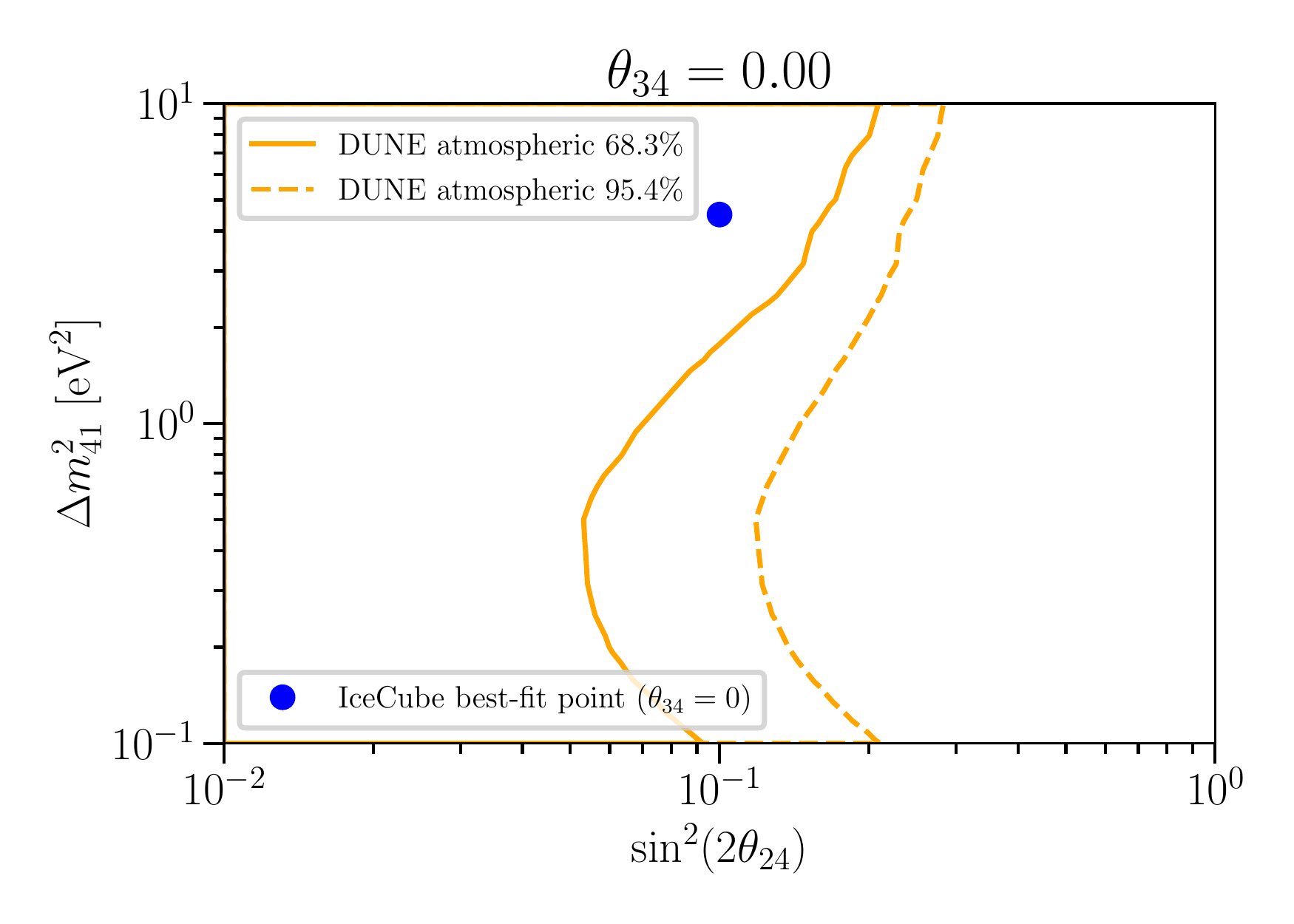}
\end{center}
    \caption{\textbf{\textit{Nominal Sensitivity to Sterile Neutrino.}}
    This plot shows the Asimov exclusion contours in the $(\Delta m^2, \sin^2 \theta_{24})$ parameter space with $\theta_{34}$ fixed to zero, assuming a 3-neutrino null hypothesis and Wilks' asymptotic approximation with three degrees of freedom.
    The IceCube best-fit point is shown as a blue circle, as well as what we expect the IceCube best-fit point to shift to (the ``test-point'') as a blue cross.
    In this region of the $\theta_{34}$ parameter space, the excluded region does not cover the IceCube best-fit point, and barely covers the test-point.
}
\label{fig:th34_0_sensitivity}
\end{figure}

\begin{figure}[tb]
\begin{center}
\includegraphics[width=\linewidth]{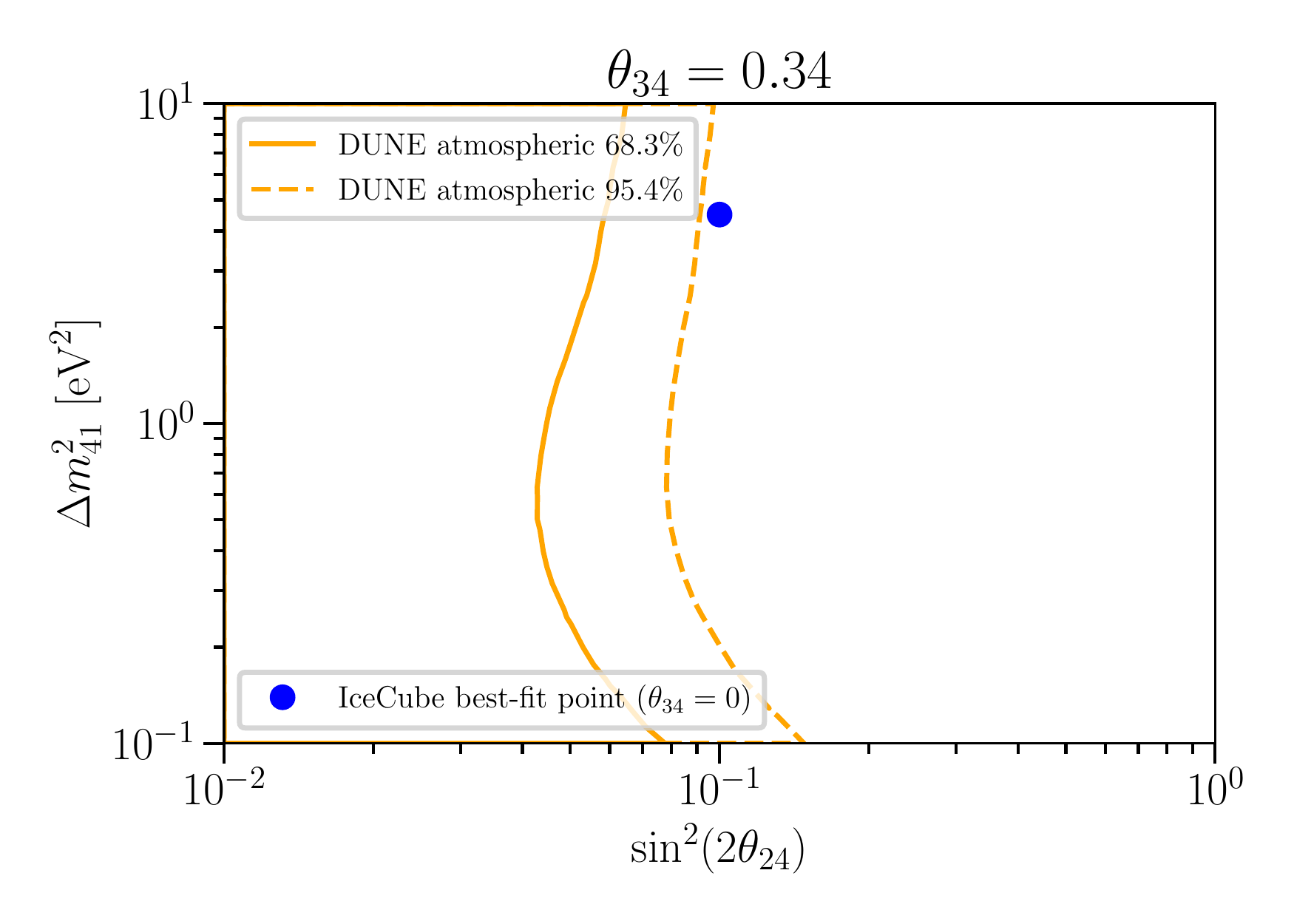}
\end{center}
\caption{\textbf{\textit{Expected Sensitivity to Sterile Neutrino.}}
    This plot shows the Asimov exclusion contours in the $(\Delta m^2, \sin^2 \theta_{24})$ parameter space with $\theta_{34}$ fixed to $0.34$, assuming a 3-neutrino null hypothesis and Wilks' asymptotic approximation with three degrees of freedom.
    In this region of the $\theta_{34}$ parameter space, the excluded region cover both the IceCube best-fit point and the test-point.
}
\label{fig:th34_34_sensitivity}
\end{figure}

For the (3+1) sterile neutrino model, the most relevant parameters for muon neutrino disappearance are $\Delta m^2$ and $\theta_{24}$.
These parameters control the location and shape of the matter resonance, and have been constrained by the IceCube muon neutrino disappearance measurements.
However, IceCube's measurements and others set $\theta_{34}$ to zero.
Non-zero $\theta_{34}$ can smear the $\bar\nu_\mu$ matter resonance to lower energies, and cause additional disappearance for $\nu_\mu$ below the resonance energy.
This modification to the oscillation signature can affect measurements of the matter resonance if not accounted for.
In the presence of true non-zero $\theta_{34}$ we expect IceCube's current measurement to be biased towards larger values of $\Delta m^2$.
Figure~\ref{fig:th34_0_sensitivity} shows the Asimov sensitivity of the 5~year scenario to the $\Delta m^2$ and $\theta_{24}$ parameters, assuming $\theta_{34}=0$.
For this value of $\theta_{34}$, the sensitivity of DUNE in the 5~year scenario does not cover the IceCube best-fit point.
However, the IceCube data constraints show a deficit in the down-going region, below the matter resonance energy, exactly where we expect additional disappearance from non-zero $\theta_{34}$.
We expect non-zero $\theta_{34}$ will be more compatible with the data.
DUNE will also have improved sensitivity for larger values of $\theta_{34}$, as shown in Fig.~\ref{fig:th34_34_sensitivity}; such that the $\SigmaTwo$ excluded region covers the IceCube best-fit point.
This means that DUNE and IceCube will be complimentary in their atmospheric sterile neutrino searches, and together may be able to differentiate between zero and non-zero values of $\theta_{34}$.

\begin{figure}[tb]
\begin{center}
\includegraphics[width=\linewidth]{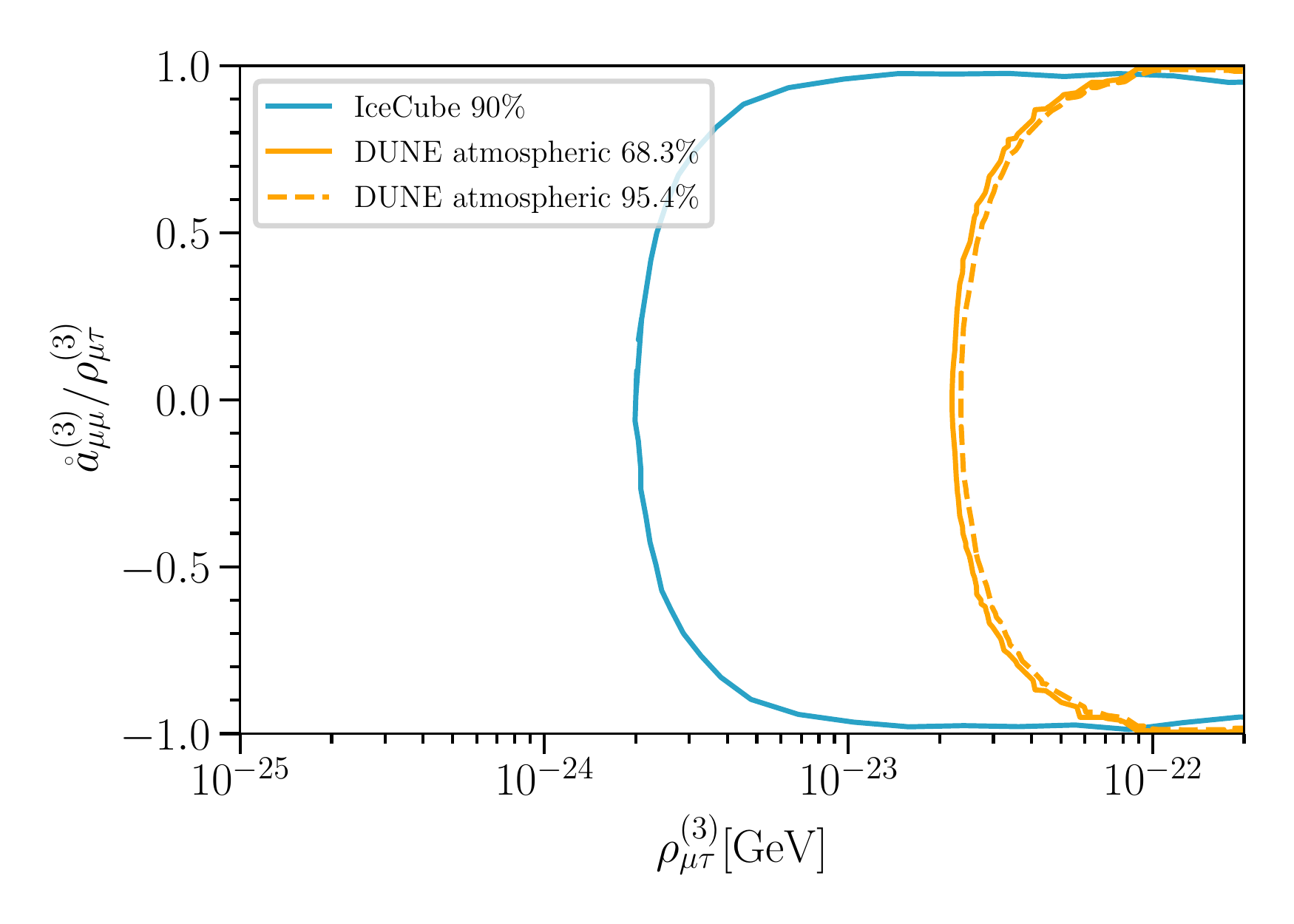}
\end{center}
\caption{\textbf{\textit{Expected Sensitivity to Lorentz Violation.}}
    This plot shows the Asimov exclusion contours in the $\rho^{(3)}_{\mu\tau}, \mathring{a}^{(3)}_{\mu\mu}/\rho^{(3)}_{\mu\tau}$ parameter space profiled over $\mu\tau$ phase and with higher dimensional operators fixed to zero.
    The null hypothesis is standard oscillation and the contours are drawn assuming Wilks' asymptotic approximation with two degrees of freedom.
}
\label{fig:lv_sensitivity}
\end{figure}

For the Lorentz violating scenario, the most relevant parameters for muon neutrino disappearance are $\rho_{\mu\tau}^{(d)}$ and $a_{\mu\tau}^{(d)}/\rho_{\mu\tau}^{(d)}$.
Additional sensitivity exists on the corresponding $\mu e$ parameters, however we expect this to be less sensitive due to the presence of the matter potential in the electron flavor. 
To obtain the Lorentz-violation sensitivities, we scan over the quantities $\rho_{\mu\tau}^{(d)}$ and $a_{\mu\mu}^{(d)}/\rho_{\mu\tau}^{(d)}$ for dimensions three. Fig.~\ref{fig:lv_sensitivity} shows the predicted sensitive region obtained for the dimension-3 coefficients in comparison with the results using high-energy neutrinos from IceCube.
The sensitivities for the maximum-flavor-violating scenario, namely $a_{\mu\mu}^{(3)} = 0$, show weaker sensitivities than the IceCube results.
However, the DUNE analysis proposed here improves over the IceCube constraints for the non-maximum-flavor-violating cases, namely near $ a_{\mu\mu}^{(3)}/\rho_{\mu\tau}^{(3)} = \pm 1 $.
Although not shown here, the differences between DUNE and IceCube are similar for higher dimensional operators.

\section{Conclusions}

In this work we propose a new way to study neutrino energies between $\SI{100}\GeV$ to $\SI{1}\TeV$ in a LarTPC by making use of the stochastic losses along the muon track. 
This energy range is uniquely accessible to upcoming large LarTPC such as DUNE, since the typical muon lengths in this regime cannot be contained inside large detectors and the energy losses cannot be properly measured in very-large-volume detectors such as IceCube due to the course spacing.
In order to estimate the sensitivity of DUNE to new physics signatures that maybe lurking in this unexplored energy range, we have developed a detailed simulation of a DUNE-like detector which we make publicly available~\cite{DUNEAtmo}.
Using our simulation we estimate the sensitivity of DUNE to two physics scenarios: light sterile neutrinos, motivated by the short-baseline anomalies, and Lorentz symmetry violation, motivated by quantum gravity and grand unifying theories.
Our results show that DUNE provides complementary measurements on sterile neutrinos when $|U_{\tau 4}| \neq 0 $ ($\theta_{34} \neq 0 $) and covers new parameter space in Lorentz violation for non-maximally-flavor violating scenarios.

\section*{Acknowledgements}

JMC and AS are supported by NSF grant PHY-1801996.
CAA and BS are supported by the Faculty of Arts and Sciences of Harvard University.

%\pagebreak
\bibliographystyle{apsrev}
\bibliography{lv_dune}

\end{document}